\documentclass[10pt,journal,compsoc]{IEEEtran}

\usepackage[T1]{fontenc}
\usepackage{amsmath,amsfonts}
\usepackage{algorithmic}
\usepackage{algorithm}
\usepackage{array}
\usepackage[caption=false,font=normalsize,labelfont=sf,textfont=sf]{subfig}
\usepackage{textcomp}
\usepackage{stfloats}
\usepackage{url}
\usepackage{romannum}
\usepackage{verbatim}
\usepackage{graphicx}
\usepackage{graphicx}
\usepackage{xcolor}
\usepackage{cite}
\usepackage{wrapfig}
\usepackage{hyperref}
\usepackage{booktabs}
\usepackage{tabularx} 
\usepackage{enumitem}
\usepackage{ulem}
\IEEEpubidadjcol
\def\oursName{ChartBlender}

\begin{document}

\title{ChartBlender: An Interactive System for Authoring and Synchronizing Visualization Charts in Video}
\author{Yi He, Yuqi Liu, Chenpu Li, Ruoyan Chen, Chuer Chen, Shengqi Dang, and Nan Cao
\thanks{Yi He, Yuqi Liu, Chenpu Li, Ruoyan Chen, Chuer Chen, Shengqi Dang, and Nan Cao are with the Intelligent Big Data Visualization Lab, Tongji University. Email: \{heyi\_11, 2333349, 2251319, 2433555, chuerchen, dangsq123, nan.cao\}@tongji.edu.cn. }
\thanks{Nan Cao is the corresponding author. Email: nan.cao@gmail.com}}

\markboth{Journal of \LaTeX\ Class Files,~Vol.~14, No.~8, August~2021}%
{Shell \MakeLowercase{\textit{et al.}}: A Sample Article Using IEEEtran.cls for IEEE Journals}

\IEEEpubid{0000--0000/00\$00.00~\copyright~2021 IEEE}
\newcommand{\hy}[1]{{\color{black}{#1}}}
\maketitle

\begin{abstract}
Embedded data visualizations have emerged as a powerful narrative medium for conveying complex information within video footage. However, creating such content remains labor-intensive, as existing workflows rely on manual frame-by-frame adjustments to ensure spatial and temporal consistency. To address these challenges, we present ChartBlender, an interactive authoring system designed to streamline the creation, embedding, and automatic synchronization of data visualizations within video scenes. We develop a tracking pipeline that supports both
object and camera tracking, ensuring robust alignment of visualizations with dynamic video content. To maintain visual clarity and aesthetic coherence, we also explore the design space of video-suited visualizations and develop a library of customizable templates optimized for video embedding. We evaluated ChartBlender through two controlled experiments and expert interviews with five domain experts. Results show that our system enables accurate synchronization and accelerates the production of data-driven videos.

\end{abstract}

\section{Introduction}

\hy{
\IEEEPARstart{E}mbedding data visualizations into videos has emerged as a powerful way to communicate complex information~\cite{idvxNVSurvey,amini2015understanding,shi2021autoclips,10891192,7539328}. This approach is now widely adopted in domains such as sports~\cite{iball}, traffic~\cite{hsieh2016traffic}, and journalism~\cite{covid19_youtube_2021}. By integrating charts into footage, it can provide timely and contextual insights while allowing viewers to stay focused on the underlying footage~\cite{iball}. Real-world examples of this visual medium can be seen in a video from \textit{The Economist}\footnote{\url{https://www.youtube.com/watch?v=pFaGbtDJFwA\&list=WL\&index=8
 (4'35''--5'20'')}} (Fig.~\ref{fig:example}(a)). In this segment, a line chart is anchored to a wall and remains synchronized with camera motion to illustrate behavioral changes during and after the pandemic.
\begin{figure}
    \centering
    \includegraphics[width=1\linewidth]{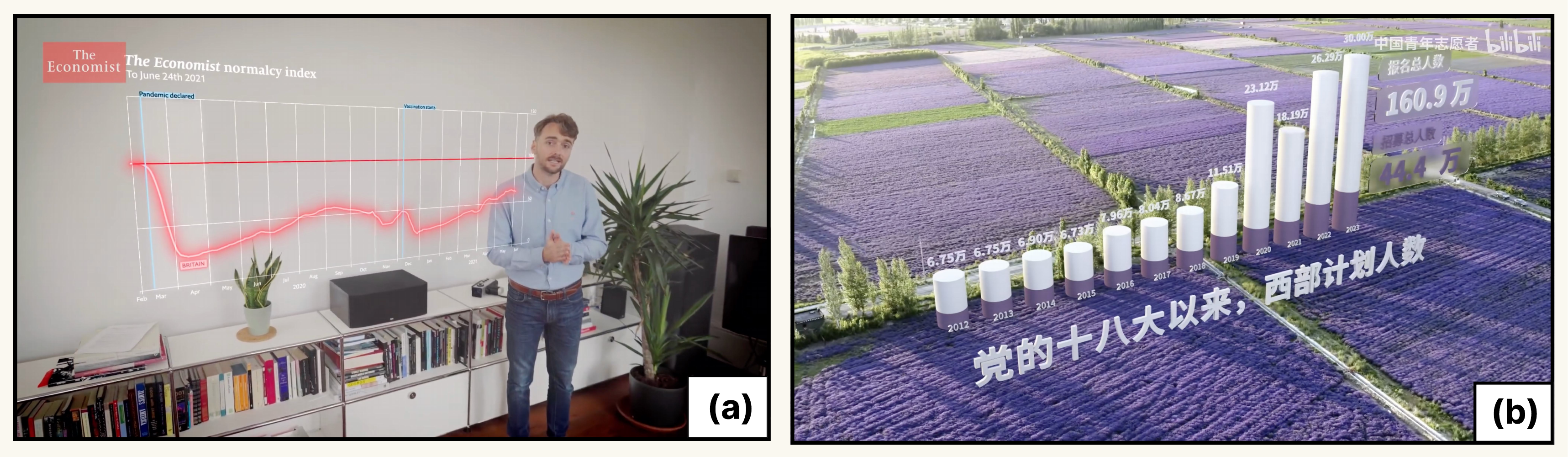}
    \caption{\hy{Real-world examples of data visualizations embedded in videos. (a) A frame from \textit{The Economist}, where a line chart illustrates behavioral changes during and after the pandemic. (b) A frame from a national media organization in China, where a bar chart shows year-by-year growth in participant numbers. }}
    \label{fig:example}
\end{figure}

Despite growing interest, producing videos with embedded visualizations remains a challenging task. Our collaboration with a national Chinese media organization~\footnote{\url{http://edu.people.com.cn/n1/2023/0915/c1006-40078200.html (2'44'', 3'25'')}} (Fig.~\ref{fig:example}~(b)) revealed critical workflow bottlenecks. During the collaboration, we were tasked with embedding a 3D bar chart into an aerial landscape shot to visualize participant growth. The chart needed to remain synchronized with camera motion. However, to the best of our knowledge, no dedicated authoring tool provides motion-aware synchronization tailored to embedded data visualizations. In practice, we relied on general-purpose video-edit software and laborious manual keyframing of the chart's pose and trajectory. Even minor revisions often triggered substantial rework, and producing a 10-second clip could take nearly a full workday. These bottlenecks directly informed our design requirements and system goals, highlighting an urgent need for an integrated, motion-aware authoring system for professional data-driven video production.
}

\hy{Within the visualization community, interest in video-embedded visualizations is also increasing. However, existing tools do not meet the requirements of motion-aware visualization authoring in real-world production. For example, sports annotation systems~\cite{7539370, iball, swimflow} demonstrate the value of integrating tracked graphics into video narratives, yet they often rely on domain-specific pipelines, fixed visual encodings, or approximate 2D overlays rather than full scene-consistent integration. Meanwhile, some AR/VR-based authoring tools~\cite{8611113,VisTellAR} enable end users to quickly place charts onto video, emphasizing ease of interaction and shareability; but they offer limited visual and stylistic control, make precise alignment and post-capture refinement difficult, and scale poorly to multi-chart compositions. Commercial video editing platforms (e.g., After Effects~\cite{adobe_ae}, Blender~\cite{blender}, DaVinci Resolve~\cite{davinci}) provide motion tracking and compositing capabilities, but visualization embedding typically requires labor-intensive workflows, substantial expertise, and disjoint processes that separate data design from in-context visual authoring. Overall, while motion tracking is widely available, existing tools rarely couple it tightly with visualization authoring, limiting dynamic, context-aware chart placement in complex scenes with editable visual encodings.}

Therefore, there is a pressing need for an interactive authoring system that supports flexible and efficient embedding of data visualizations into complex video scenes while maintaining accurate motion synchronization throughout the video.
However, to design such an effective system, three challenges need to be tackled: (1) robustly tracking motion by accounting for both object movement and camera pose variation to ensure consistent synchronization between the visualization and the scene; (2) supporting generalized chart designs that remain legible and informative under dynamic video conditions; and (3) enabling simple, integrated interactions for user-driven authoring and iterative refinement. Addressing these challenges is essential for precise and seamless visualization embedding in real-world videos.

To address these challenges, we present \textit{\oursName}, an authoring system that embeds and synchronizes data visualizations with video content. Users specify a tracking point and choose a tracking mode, and the system automatically generates the motion trajectory. At any time, users can adjust both the chart’s visual style and its 3D position in the scene. In this way, the system supports the human-machine collaboration to generate data-driven videos rapidly while ensuring the visualization appears visually coherent and seamlessly embedded into the video. The major contributions of the paper are as follows.

\begin{itemize}

 \item \textbf{System.} We introduce \textit{\oursName}, a novel authoring system that enables the authoring and automatic synchronization of data visualizations with video content. It features an interactive interface that supports the end-to-end creation of visualization-embedded videos through an integrated authoring workflow. We also developed an algorithm pipeline to support robust visualization embedding for the system.

 \item \textbf{Design Space.} We characterize the design space of visualizations suitable for video embedding. Specifically, we identify two primary motion modes, \textbf{camera tracking} and \textbf{object tracking}, and identify five commonly used chart types. Based on this analysis, we develop a library of generalized visualization templates designed to convey complex data clearly and effectively within dynamic video contexts.

\hy{\item \textbf{Evaluation.} We demonstrate the utility of \textit{\oursName} through a set of example videos created from real-world data, two controlled experiments that validate the tracking pipeline, and a series of interviews with five expert users.}

\end{itemize}

\section{Related Work}
\hy{In this section, we review the recent studies that are most relevant to our work, including embedding visualizations in video, video editing software, and point tracking.}

\hy{\subsection{Embedding Visualizations in Video}
Embedding visualization, as a form of embedded data representation, places data-encoded graphics in close spatial correspondence with real-world referents (e.g., objects or locations), enabling viewers to interpret data in context rather than consulting a separate chart view \cite{7539328}. When the target medium is video, this coupling should be preserved under continuous camera and scene motion, making the problem a form of visualization in motion\cite{yao2020}. In this setting, relative motion and limited viewing time can affect how accurately and efficiently viewers read values, while professional production further requires stable spatiotemporal alignment and reliable legibility under changing viewpoints, cluttered backgrounds, and occlusions.

Early efforts such as SmartShot~\cite{smartshot} examined static embedding by optimizing chart placement to reduce occlusion of salient content and to balance aesthetic integration with readability. Beyond appearance, however, realistic embedding in dynamic 3D footage requires viewpoint-stable registration. As a result, perspective consistency, correct depth ordering, and occlusion handling become essential. To preserve visual coherence, embedded charts should follow camera trajectories and object dynamics; accordingly, a range of techniques and systems have been proposed~\cite{7539370,10.1145/3411764.3445337,10.1145/3411764.3445344,10.1109/TVCG.2021.3114774}. For example, SwimFlow~\cite{swimflow} embeds animated visualizations in swimming videos by annotating key positions and interpolating trajectories, which works well for predictable motion but is less general for arbitrary scenes. iBall~\cite{iball} overlays dynamic cues in basketball videos and adapts displays based on gaze; however, under camera motion the graphics are only approximately anchored and do not provide perspective-consistent 3D registration. Overall, many existing systems rely primarily on 2D overlays and tracking, offer limited depth-aware occlusion reasoning, and are tailored to specific domains.

More recent work has explored general-purpose authoring systems that synchronize charts with arbitrary videos. With advances in AR/VR, it has become easier to create motion-following overlays. Chen et al.~\cite{8611113} introduced MARVisT, a mobile authoring tool that lets non-experts bind data to virtual glyphs and real objects, and systems such as VisTellAR~\cite{VisTellAR} enable casual users to overlay visualizations on videos quickly. However, these end-user tools typically trade expressive control and production robustness for accessibility: they provide limited chart grammars and styling controls, make precise alignment and post-capture refinement difficult, and scale poorly to multi-chart compositions.

In this work, we introduce \oursName\ as an interactive authoring system for embedding and synchronizing charts in video. \oursName\ supports 3D object tracking and perspective-consistent registration to anchor charts stably in dynamic scenes. Unlike prior systems that rely on fixed overlays or narrow workflows, \oursName\ enables flexible, in-context authoring: designers can edit chart types, styles, and data mappings directly within the video while maintaining precise control over placement and timing. This integration streamlines the creation of motion-aware embedded visualizations for professional production.}

\hy{
\subsection{Video Editing Software}
Nowadays, modern video editing and VFX tools provide motion tracking capabilities that allow designers to attach graphics or objects to dynamic scenes. After Effects~\cite{adobe_ae}, for example, includes point tracking, spline-based tracking, and 3D camera tracking, enabling powerful compositing; in practice, however, achieving stable, perspective-consistent overlays, especially in the presence of fast motion or occlusions, often requires substantial manual refinement. Blender~\cite{blender} provides camera tracking and 3D scene integration within a full 3D authoring environment, but producing polished results typically requires expertise and nontrivial setup. Final Cut Pro~\cite{final_cut} offers an accessible, machine-learning-based object tracker for effects such as blurring or target-following, which is efficient for lightweight editing but provides less fine-grained control. DaVinci Resolve’s Fusion module~\cite{davinci} supports planar tracking and 3D camera tracking in a node-based workflow, yet its learning curve and context switching between editing and compositing can hinder iterative authoring. Foundry Nuke~\cite{nuke} is an industry-standard compositor with advanced planar tracking and 3D camera solving; it offers high precision but is costly and complex, and is therefore most commonly used in professional studio pipelines.

Across these tools, automatic tracking alone often falls short for chart-embedded video authoring because (1) tracking is not visualization-specific and frequently demands frame-by-frame refinement for chart stability, (2) professional workflows impose steep learning curves and high authoring overhead, and (3) chart design and tracking are decoupled, making in-context preview and iterative refinement of visualization appearance cumbersome. By contrast, \textit{\oursName} offers a lightweight and learnable workflow with interactive controls that allow authors to adjust chart styling and placement while immediately previewing the results in context.}

\subsection{Point Tracking}
Tracking arbitrary points in videos has witnessed substantial progress in recent years~\cite{zhou2020tracking,gauglitz2011evaluation,10.1007/978-3-030-58548-8_28}, with methods such as TAP-Net \cite{hu2020tap} demonstrating the ability to reliably follow both static and dynamic points throughout a video sequence. These approaches typically leverage appearance-based cues and temporal consistency to maintain accurate correspondences, even under significant motion and occlusion. However, focusing solely on 2D tracking is often insufficient in practical applications. While a dynamic point in the scene can be tracked frame-by-frame, certain use cases, such as visual overlays or interactive annotations, require this point to remain anchored relative to the static scene background, rather than to moving objects or transient visual cues. To achieve this, it becomes necessary to disentangle camera motion from scene dynamics, which calls for robust camera pose estimation.

Fortunately, a variety of mature techniques now exist for estimating camera pose from monocular or multi-view video streams~\cite{xu2024critical, guan2024survey}, ranging from classical SLAM-based pipelines~\cite{campos2021orb, vidal2018ultimate,7219438} to deep learning methods\cite{shavit2019introduction,schonberger2016structure, Hidalgo-Carrio_2022_CVPR,zhou2017unsupervised,en2018rpnet}. The integration of 2D point tracking with camera pose estimation enables a more unified paradigm: tracking arbitrary points in 3D space. This formulation has recently attracted increasing attention~\cite{koppula2024tapvid,yu2023videodoodles,xiao2024spatialtracker}, as researchers extend traditional 2D tracking frameworks into the 3D domain.
For instance, SpatialTracker\cite{xiao2024spatialtracker} is the first work specifically designed for the task of 3D point tracking. It estimates the 3D trajectory of arbitrary pixels by combining image features with monocular depth estimation, and propagates information through a graph neural network (GNN) to model long-range spatiotemporal motion. Similarly, TAPIR-3D\cite{koppula2024tapvid} is an experimental extension of the original 2D point tracking model TAPIR, designed to output 3D point trajectories. It initializes keypoints in each frame, refines their 2D tracks using temporal context from adjacent frames, and subsequently lifts them to 3D space using depth estimation models such as ZoeDepth~\cite{bhat2023zoedepth}.

We propose a pipeline that enhances the accuracy of 3D point tracking by integrating it with the practical requirements of embedding visualizations into real-world video content. Our system enables users to anchor visualizations to fixed locations within the scene or dynamically attach them to moving objects, regardless of camera motion or scene complexity. This flexibility allows for robust visualization embedding across a wide range of video scenarios.

\hy{
\section{System Design and Overview}
In this section, we present the design of \oursName. We first introduce our collaboration with a national Chinese media organization. Drawing on their feedback and our observations, we distill four design requirements (R1–R4) and then provide an overview of the system.

\subsection{Collaboration with Professional Data Journalists}
The development of \oursName\ was motivated by a collaboration with a national Chinese media organization. Our research team was invited to provide technical support for a micro-documentary on the College Student Volunteer Service Western Project, which aimed to narrate volunteers’ stories and the program’s long-term impact on regional development. We assembled a project team of two professors and four graduate students with expertise in data visualization. We worked closely with five professional data journalists from the organization (average 15 years of work experience), who were highly familiar with data videos. Using raw footage and data provided by the organization, we were tasked with producing three chart-embedded video segments that required high-precision visual integration. One segment embedded a 3D bar chart into an aerial drone shot to depict year-by-year changes in volunteer participation, with the chart appearing anchored to the terrain. The other two segments overlaid key indicators that had to remain tightly aligned with moving targets in the footage.

The collaboration followed a 15-day production cycle with daily online review meetings and rapid iterative revisions. In each meeting, our team presented the latest composites together with intermediate project files, and the journalists evaluated them shot by shot. They requested targeted adjustments such as changing anchoring targets, fine-tuning entrance and exit timing, refining visual styling for on-screen legibility, and repositioning overlays. Between meetings, we implemented these requests and circulated updated renders for asynchronous feedback, which was then consolidated and addressed in subsequent reviews. This recurring loop of authoring, critique, and revision not only ensured visual precision but also exposed where existing workflows break down, what corrections experts repeatedly perform, and which controls they consider essential for professional-grade results.

To support the task, we surveyed prior systems for chart-embedded video authoring. One notable system is VisTellAR~\cite{VisTellAR}, which enables end users to capture short videos and embed data visualizations through augmented reality (AR). VisTellAR tracks static elements in the scene and anchors charts accordingly, offering a lightweight, mobile-friendly workflow. Although VisTellAR is not open source, we introduced it to our collaborators as a concrete reference for discussing chart-in-video authoring. They noted, however, that unlike end-user workflows, professional data journalists (our target users) typically embed charts during post-production and demand substantially higher precision and finer-grained control. Across production iterations and review discussions, three needs repeatedly surfaced: (1) accurate motion tracking that supports both static structures and dynamic objects; (2) compatibility with diverse footage, including videos without camera metadata, which is common in practice; and (3) support for multiple dynamic charts within a single video, with fine-grained control over timing and placement to maintain visual coherence.

In the absence of dedicated tools, we followed a typical professional workflow to produce the segments. We first created charts in 3D software, exported them as 3D assets, imported the assets into Adobe After Effects, applied automatic motion tracking to obtain an initial alignment, and then refined the results by adjusting position and orientation frame by frame before final compositing. During this process, we observed several bottlenecks that hinder seamless and efficient embedding. Although professional compositing software provides tracking functions, tracking results can be unstable in challenging shots, requiring repeated adjustments of position, scale, and orientation to maintain spatiotemporal consistency. Moreover, chart authoring and motion tracking are decoupled, preventing designers from previewing and refining charts in context while editing. In our experience, producing a 10-second clip with this workflow can take several hours, even a full workday, with much of the effort devoted to repetitive refinements rather than higher-level design work. These observations motivate a dedicated authoring tool that tightly integrates chart design with robust tracking to automate chart–video synchronization while preserving the precise control required in professional production.
\subsection{Design Requirements}
Through discussions with our collaborators, we distilled four design requirements~(R1-R4) that guided the development of \oursName, which shaped both the system's architecture and interface.
}
\begin{itemize}
    \item[{\textbf{R1}}]\textbf{Video-Suited Visualization Design.} Visualizations should be designed to present data clearly while preserving the visual coherence of the scene, minimizing interference with the underlying video content.
 
    \item[{\textbf{R2}}]\textbf{Depth-Aware Embedding.} Visualizations should be embedded into the video by aligning with camera depth and adopting a perspective consistent with the video scene. Such alignment ensures that the visualizations appear seamless to the scene, rather than as artificial post-production overlays.

    \item[{\textbf{R3}}]\textbf{Motion Synchronization.} As the video plays, the visualizations should move in synchrony with the video, including the movement of both the camera and objects in the scene. To achieve this synchronization, it is necessary to accurately track both camera and object motion.

     \item[{\textbf{R4}}]\textbf{Coordinated Multi-Chart Embedding.} The system should support the integration of multiple visualizations within a single video, allowing authors to coordinate their spatial placement and temporal sequencing. This enables richer data storytelling and ensures that each chart appears at the appropriate moment and location in the visual narrative.

\end{itemize}
\begin{figure}
    \centering
    \includegraphics[width=1\linewidth]{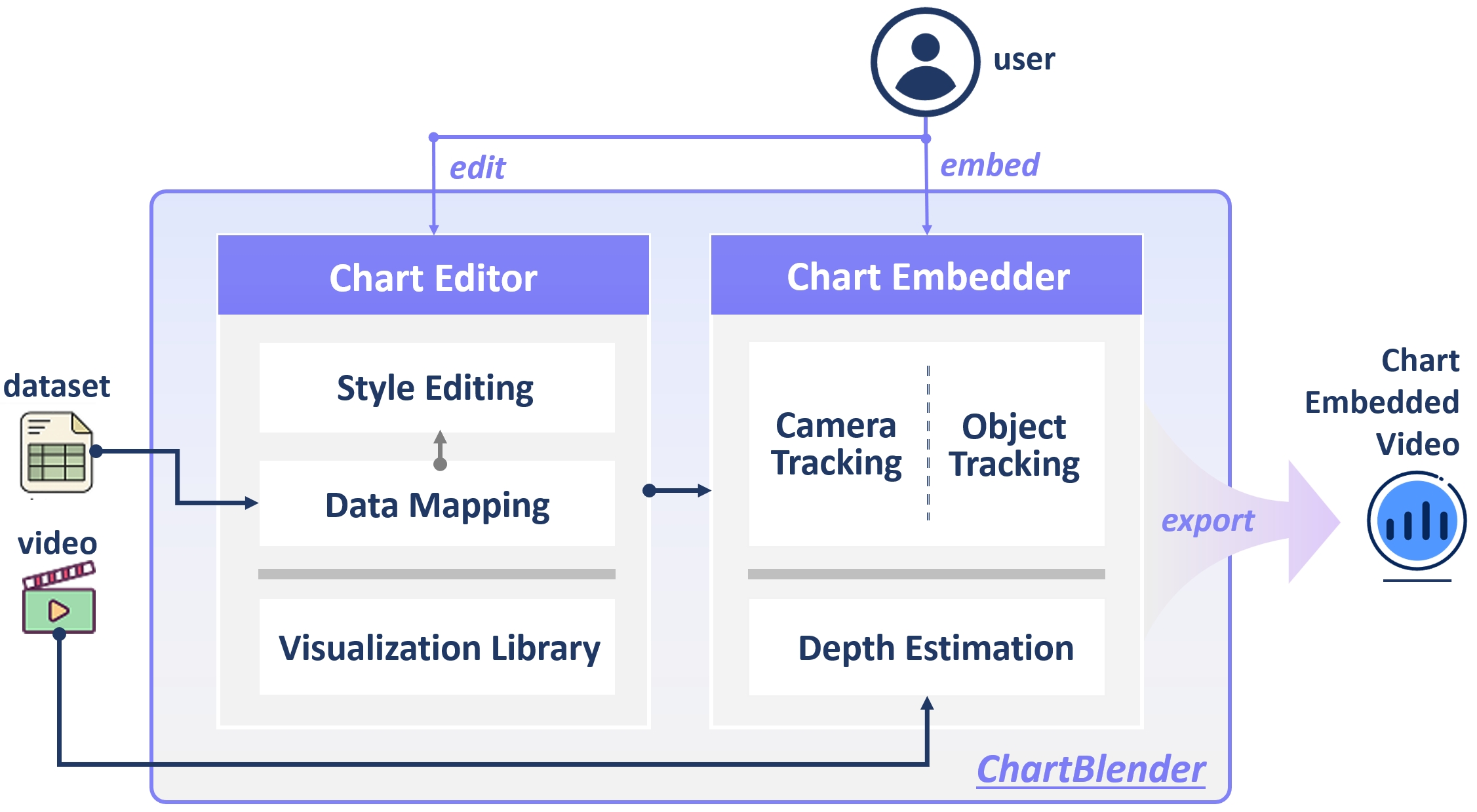}
    \caption{\textit{\oursName} system consists of two modules: Chart Editor and Chart Embedder.}
    \label{fig:workflow}
\end{figure}
To fulfill these requirements, we developed the \oursName\ system. Fig.~\ref{fig:workflow} illustrates the system architecture and the embedding pipeline. The system consists of two main modules: (1) the Chart Editor, and (2) the Chart Embedder.
Generally, a user starts creating a visualization-embedded video by uploading a dataset and interactively configuring the visualization using the Chart Editor. This module allows users to customize various properties such as data mapping and style editing\textbf{ (R1)}. Afterward, users have a real-time preview of the chart's 3d placement and appearance within the video to the Chart Embedder \textbf{(R2)}.
The Chart Embedder incorporates a tracking algorithm that synchronizes visualizations with motion in the video and operates without requiring any camera metadata. It first performs depth estimation on the uploaded video, then anchors the visualization to the scene based on the user's selected tracking mode and target \textbf{(R3)}. Finally, users can further adjust the chart, trim the timeline, and incorporate additional elements to produce smoother, more engaging, visualization-enhanced videos~\textbf{(R4)}.
Collectively, these modules provide a lightweight yet powerful interactive workflow that automates chart embedding, reduces manual effort, and lowers the barrier for users. At the same time, the system supports the creation of visually coherent visualizations that integrate seamlessly into dynamic video scenes. While the Chart Editor functions as a relatively standard visualization creation module, in the following section, we focus on the Chart Embedder, which constitutes the core contribution of our system.
\section{Chart Embedder}
\begin{figure*}
    \centering
    \includegraphics[width=1.0\linewidth]{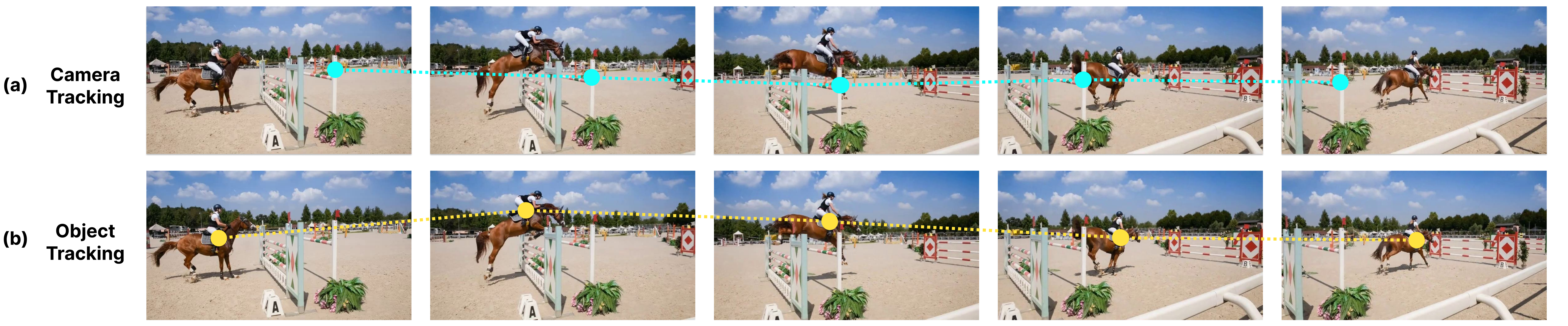}
 \caption{An illustration of the two motion embedding modes supported by our system: \textit{Camera Tracking} and \textit{Object Tracking}. In the \textit{Camera Tracking} mode, the visualization remains anchored to a fixed point in the scene, attaching a chart to a stationary fence in the video. In contrast, \textit{Object Tracking} binds the visualization to a moving object, attaching a chart to a galloping horse.}
    \label{fig:two_motions}
\end{figure*}

\begin{figure}[htbp]
    \centering
    \includegraphics[width=1\linewidth]{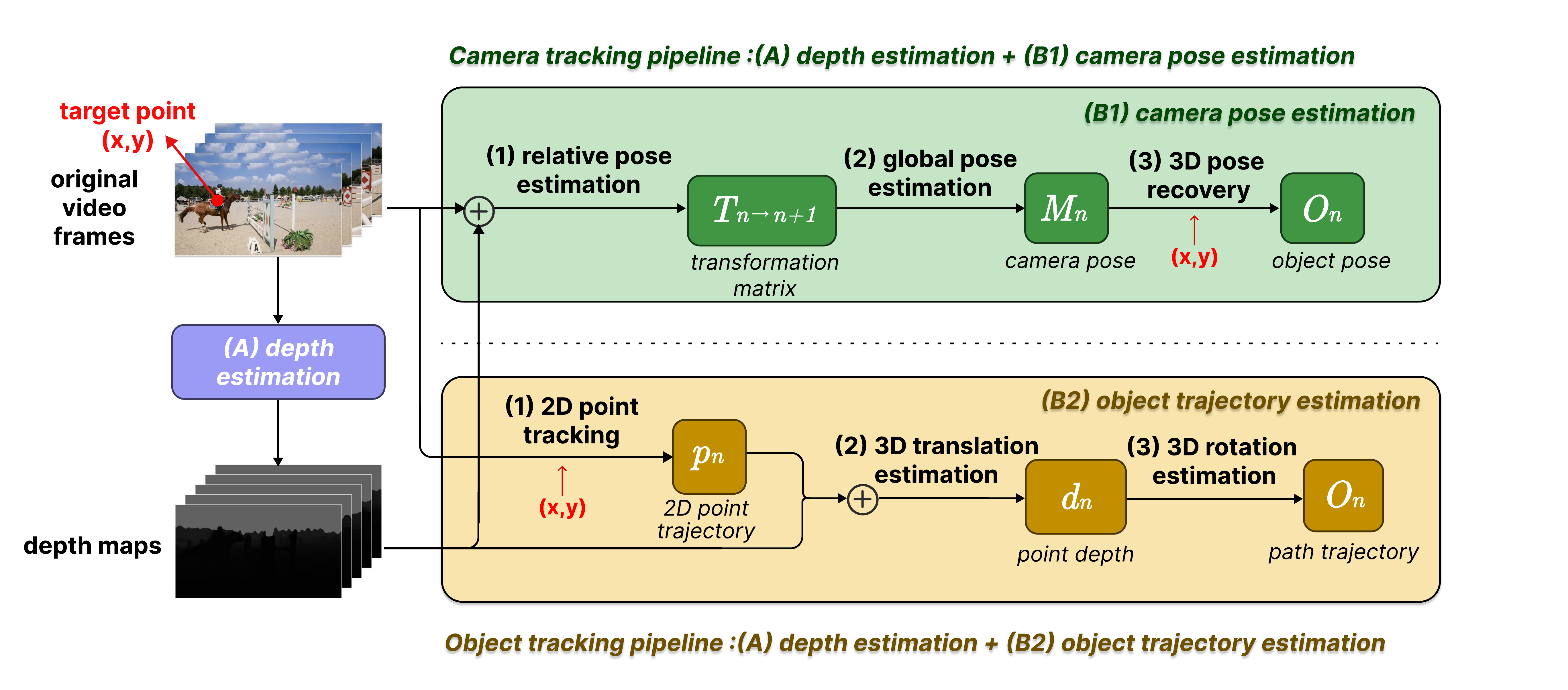}
    \caption{\hy{The two pipelines of the Chart Embedder. The \textit{camera tracking} pipeline consists of~(A) depth estimation followed by~(B1) camera pose estimation. The \textit{object tracking} pipeline consists of~(A) depth estimation followed by~(B2) object trajectory estimation. }}
    \label{fig:pipeline}
\end{figure}

\hy{The Chart Embedder module supports two tracking pipelines: \textit{camera tracking} and \textit{object tracking}. Camera tracking keeps an embedded 3D visualization spatially consistent with a static scene element throughout the video (Fig.~\ref{fig:two_motions}(a)), whereas object tracking keeps the visualization aligned with a moving target (Fig.~\ref{fig:two_motions}(b)). 

As shown in Fig.~\ref{fig:pipeline}, both pipelines start with (A) \textit{depth estimation}, which provides per-frame depth cues needed to lift 2D image measurements into 3D and to render charts with perspective-consistent projection. They then differ in their motion estimation stage: (B1) \textit{camera pose estimation} for camera tracking recovers the camera trajectory so a chart can remain anchored to a fixed location in the scene despite camera motion, whereas (B2) \textit{object trajectory estimation} for object tracking recovers the 3D motion of a target so a chart can follow the object over time.

}

\subsection{Depth Estimation}
\hy{
Unlike general-purpose 2D trackers that return image-plane trajectories, chart embedding requires 3D, scene-referenced motion signals: we need depth to lift user-selected points from pixels into 3D space and to compute poses that keep a chart perspective-consistent and spatially stable under camera motion. To obtain the required spatial context, our system uses a monocular depth estimation model, \textit{Metric3D}~\cite{yin2023metric}, to predict metric depth maps for each video frame. We adopt Metric3D for two reasons: (1) it predicts absolute metric depth without requiring ground-truth depth or camera intrinsics, enabling zero-shot deployment on in-the-wild videos; and (2) it generalizes well across diverse scenes due to large-scale training data.

Given these benefits, we apply Metric3D to each of the $N$ frames to obtain a sequence of dense depth maps ${D_n}_{n=0}^{N-1}$. For a user-specified 2D point $(x, y)$, we extract the depth value in the first frame as $d = D_0(x, y)$. This depth signal provides the foundation for subsequent camera and object tracking, enabling spatial reasoning in both modules.

\subsection{Camera Pose Estimation}
In \oursName, camera pose estimation aims to recover a consistent camera trajectory across frames so that a visualization anchored to a static scene location remains stable under camera motion. We decompose this problem into three steps.
First, we perform \textbf{(1) relative pose estimation} to compute the geometric transformation between adjacent frames, capturing local camera motion that can be robustly inferred from short temporal baselines. Second, we perform \textbf{(2) global pose estimation} by accumulating these relative transformations to obtain a per-frame camera pose in a shared coordinate system, which is required for world-consistent anchoring over time. Finally, we perform \textbf{(3) 3D pose recovery} to lift the user-specified 2D anchor (with predicted depth) into a world-space reference and derive the corresponding per-frame chart pose, enabling accurate 3D placement and projection throughout the video.}

\textbf{Relative pose estimation.} To understand how the camera moves from frame to frame, we should estimate the relative transformation between each pair of consecutive video frames. These pairwise transformations form the foundation for reconstructing the full camera trajectory.
We employ an RGBD odometry model provided by Open3D~\cite{Zhou2018,park2017colored} to compute the relative camera motion between adjacent RGBD frames. Given two consecutive frames $n$ and $n+1$, RGBD odometry estimates a transformation matrix. \hy{Specifically, we extract a transformation matrix \( T_{n \rightarrow n+1} \) that aligns the RGB and depth images. 
Each transformation composed of a rotation matrix \( R_{n \rightarrow n+1} \) and a translation vector \( t_{n \rightarrow n+1} \).}

\textbf{Global pose estimation.} While relative poses describe local motion between frames, embedding visualizations in a stable, world-aligned coordinate system requires each frame's absolute camera pose. This enables a consistent rendering of the 3D object throughout the video sequence.
\hy{We compute the absolute camera pose $M_n$ at each frame $n$ by chaining all previous relative transformations starting from the first frame: $M_n = T_{1 \rightarrow 2} T_{2 \rightarrow 3} \cdots T_{{n-1} \rightarrow n}$
This accumulation step reconstructs the global camera trajectory, allowing us to express each frame's viewpoint in the same coordinate system.}

\textbf{3D pose recovery.} 
To render the 3D visualization as if it were physically anchored in the scene, we should compute its position in the global world coordinate system. Given a user-specified 2D point \((x, y)\) in the image and its predicted corresponding depth \(d\) predicted by the \textbf{depth estimation} module, we first reproject the point back into the 3D camera coordinate system using the inverse of the intrinsic matrix of the camera \(K^{-1}\). 
Next, we transform this point into the world coordinate system using the inverse of the absolute camera pose \(M_n\), computed from accumulated relative poses. 
This transformation places the 3D visualization at a consistent location in the world coordinate system, regardless of camera motion. It ensures that the visualization remains spatially stable and visually aligned with the static background in the video.

\hy{

\subsection{Object Trajectory Estimation}
Object trajectory estimation aims to recover a per-frame 3D pose for a moving target so that an attached visualization follows the object with stable position and coherent orientation over time. We decompose this problem into three stages. Specifically, we estimate per-frame object poses via \textbf{(1) 2D point tracking}, which extracts a pixel trajectory; \textbf{(2) 3D translation estimation}, which combines the tracked 2D positions with the predicted depth to recover the object’s 3D displacement; and \textbf{(3) 3D rotation estimation}, which infers a smooth orientation change to keep the visualization consistently aligned with the object’s motion. Together, these stages produce per-frame poses that drive motion-synchronized chart embedding for dynamic targets.

\textbf{2D point tracking.} We identify salient points on the target object and track their 2D locations across frames. We use BootsTAP~\cite{doersch2024bootstap}, a high-precision keypoint tracker based on dense visual correspondence. For a target point $(x, y)$, the tracker outputs a pixel-level trajectory $p_n = (u_n, v_n)$ across frames. These trajectories capture apparent motion in image space and serve as the basis for recovering 3D displacement.

\textbf{3D translation estimation.} We lift 2D trajectories into 3D by combining them with per-frame depth. For each frame, we obtain the depth at the tracked pixel from the corresponding depth map, $d_n = D_n(p_n)$, and back-project $p_n$ to a 3D point $P_n$ in camera coordinates. Because monocular depth can be noisy and temporally inconsistent (e.g., due to low-texture regions, occlusion, or lighting changes), the recovered 3D trajectory may exhibit jitter. We therefore smooth the depth signal over time using quadratic regression within a sliding window, and then compute the per-frame translation vector $t_n$ via back-projection with the smoothed depth and intrinsic matrix $K$.

\textbf{3D rotation estimation.} We estimate object rotation from the direction of motion in 3D so that the visualization remains properly oriented as the object moves. We define the per-frame motion vector as $V_n = P_{n+1} - P_n$, and obtain a rotation $R_n$ by constraining the object’s local $x$-axis to align with $V_n$. To stabilize rotation, we first regularize 3D point trajectories by minimizing sudden changes in displacement using a second-order temporal smoothing objective:
\begin{equation}
\min_{\hat{P}} \sum_n \| \hat{P}_{n+1} - \hat{P}_n \|^2 + \lambda \sum_n \| \hat{P}_{n+2} - 2\hat{P}_{n+1} + \hat{P}_n \|^2
\end{equation}
where $\lambda$ controls the strength of the acceleration penalty. We apply the same principle in the velocity domain to encourage smooth motion direction over time, mitigating jitter and improving temporal coherence. Given the estimated translation $t_n$ and rotation $R_n$, the per-frame object pose is defined as $O_n = \{ R_n \mid t_n \}$.

}

\section{ChartBlender}
\begin{figure}
    \centering
    \includegraphics[width=1\linewidth]{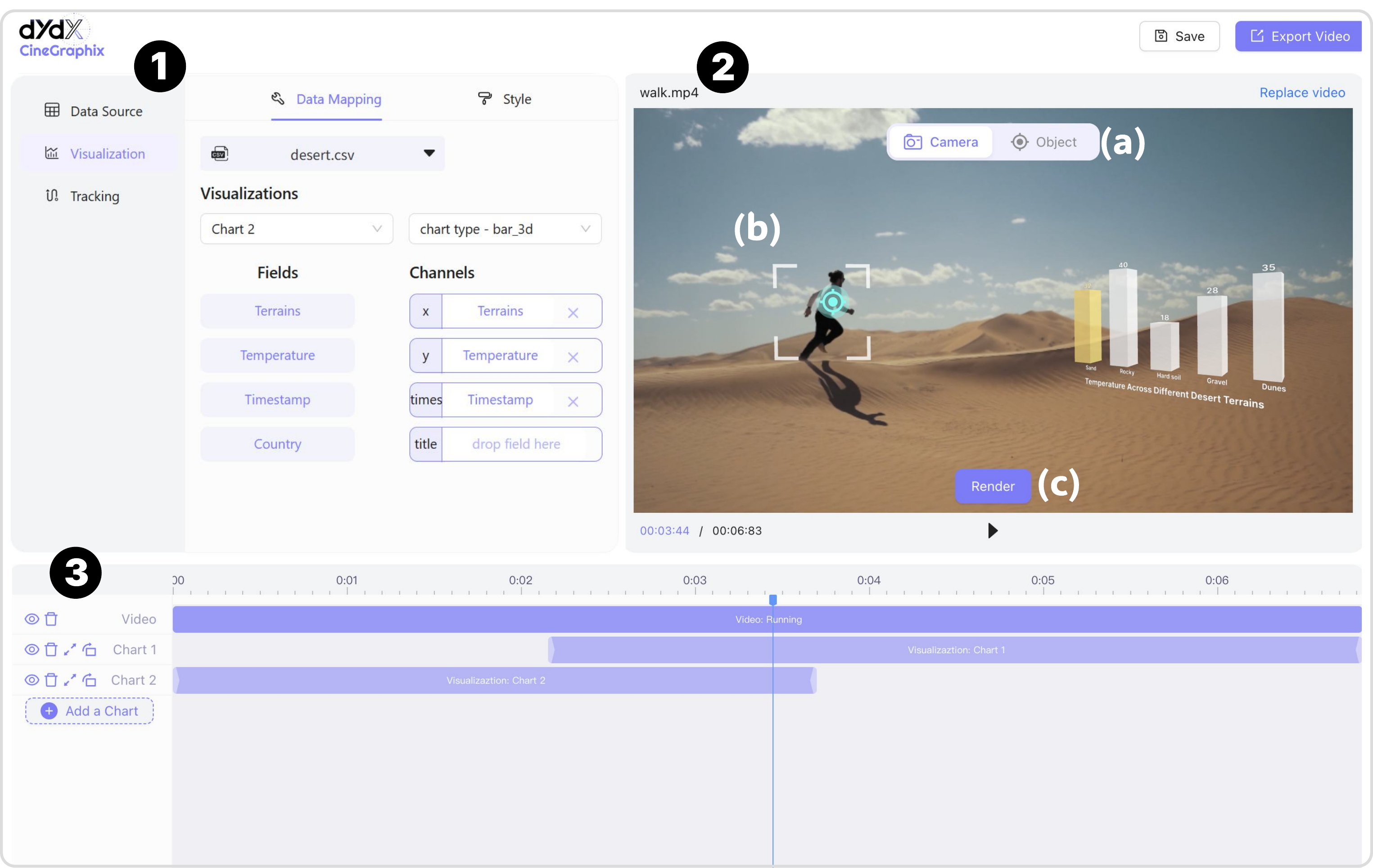}
    \caption{The interface consists of three main components: (1) the Chart View for creating and editing visualizations, (2) the Video View for embedding charts and previewing the embedding result, and (3) the Timeline View for adding, removing, and trimming chart segments.}
    \label{fig:interface}
\end{figure}
In this section, we introduce the interface design, visualization design, and interaction.

\subsection{User Interface}
The user interface, as shown in Fig. \ref{fig:interface}, consists of three major views: \textbf{ the Chart View} (Fig. \ref{fig:interface}-1), \textbf{the Video View} (Fig. \ref{fig:interface}-2), and \textbf{the Timeline View} (Fig. \ref{fig:interface}-3).
In the Chart View, users can upload a dataset, after which the raw data is displayed directly within the view. Users can then choose a chart type (e.g., \textbf{bar chart, line chart, area chart, and text}) and define encoding mappings by dragging data fields to visual channels such as the \textbf{x-axis, y-axis, timestamp, or title}. To refine the visual presentation, users can adjust stylistic properties including line thickness, color, and theme~\textbf{(R1)}.
The Video View provides a real-time preview of the current video frame. Users can manipulate the visualization’s position, rotation, and scale in 3D space, enabling precise spatial embedding within the scenes~\textbf{(R2)}. To synchronize visualizations with video motion, users first select a motion mode, either \textbf{camera tracking} or \textbf{object tracking} (Fig.\ref{fig:interface}-2(a)), and then specify target points in the video as tracking anchors (Fig.\ref{fig:interface}-2(b)). After configuration, clicking the Render button (Fig.\ref{fig:interface}-2(c)) triggers the system to compute the visualization’s motion path and composite it into the video~\textbf{(R3)}. Once rendering is complete, users can click the \textbf{Play} button to preview the embedded visualization in motion, ensuring alignment and visual consistency.

The Timeline View serves as an editing panel that supports temporal coordination and layer management. On the left side of the interface, users can add, delete, or lock chart layers to organize content across multiple visualization tracks. On the right side, a \textbf{Timeline} allows users to control the duration and timing of each chart via intuitive drag, stretch, and trim interactions~\textbf{(R4)}.
Finally, by clicking the \textbf{Export} button, users can generate the composite video with the embedded visualizations synchronized to the video.

\subsection{Visualization Design}
\begin{figure}
    \centering
    \includegraphics[width=0.9\linewidth]{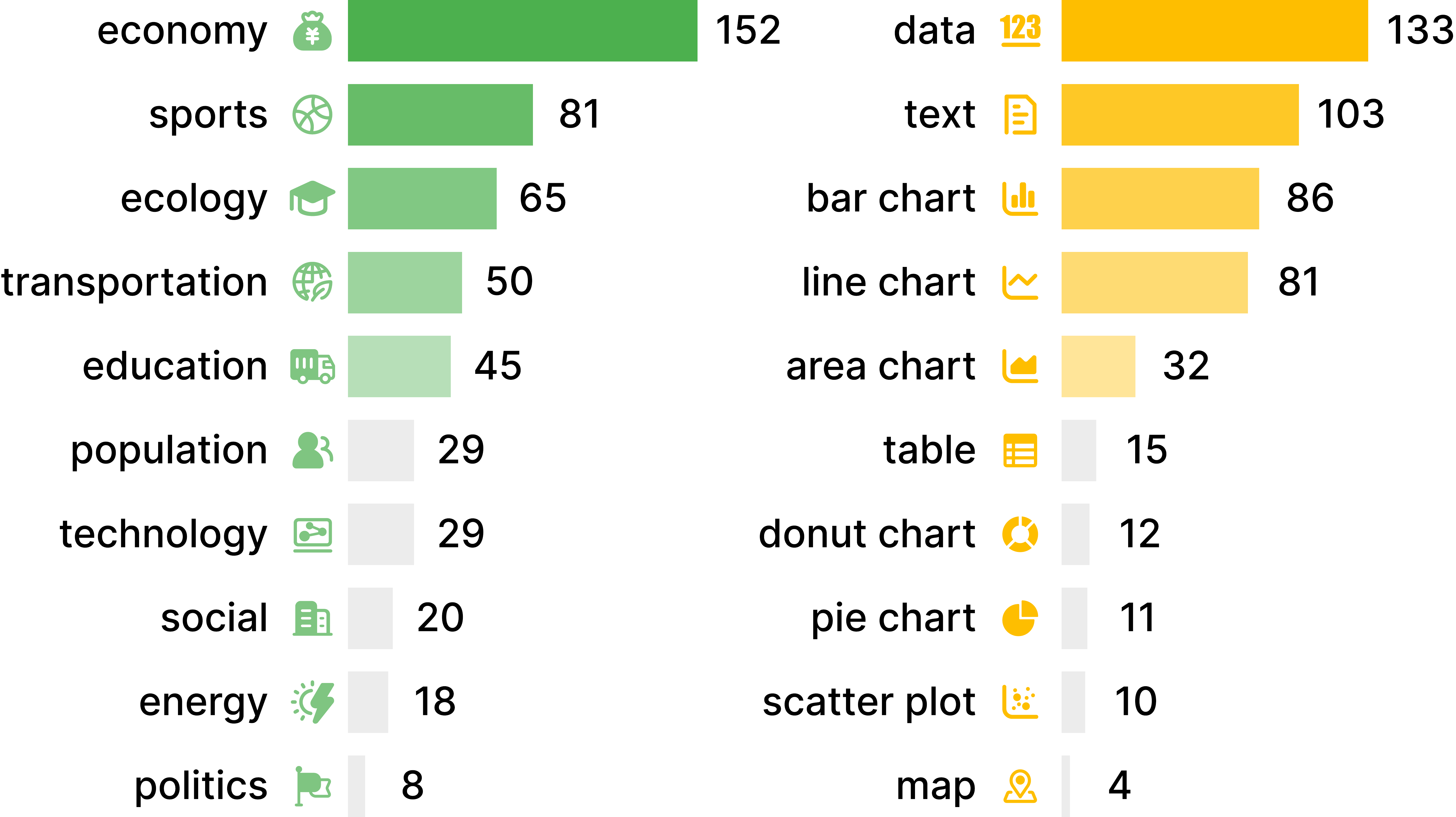} 
    \caption{Distribution of chart types and video domains.}
    \label{fig:chart}
\end{figure}

\hy{We observed that visualizations appearing in videos often differ from conventional charts. To understand how embedded visualizations are designed in practice and to inform the design of our chart library, we followed a systematic workflow: we collected a real-world video corpus and synthesized recurring strategies into a design space of embedded-visualization principles, offering practical implications for both researchers and designers.

\subsubsection{Methods}
To ground our design in practice, we assembled a corpus of 200 publicly available, high-quality videos featuring embedded visualizations from online platforms (e.g., YouTube and Bilibili) and official outlets in domains such as news, sports, technology, and education.
Each video was manually segmented at visualization entry and exit points, yielding 502 clips containing at least one embedded visualization. Two trained annotators coded each clip using a codebook that combined deductive categories (chart type, 2D/3D representation, spatial relation to the scene, motion coupling to camera/object, occlusion handling, and labeling/styling) with inductive categories for emergent patterns. After an initial calibration phase, disagreements were resolved through discussion. We also recorded domain metadata spanning ten categories (e.g., economy, sports, ecology, and transportation), as summarized in Fig.~\ref{fig:chart}.}
\begin{figure*}[t]
    \centering
    \includegraphics[width=1\linewidth]{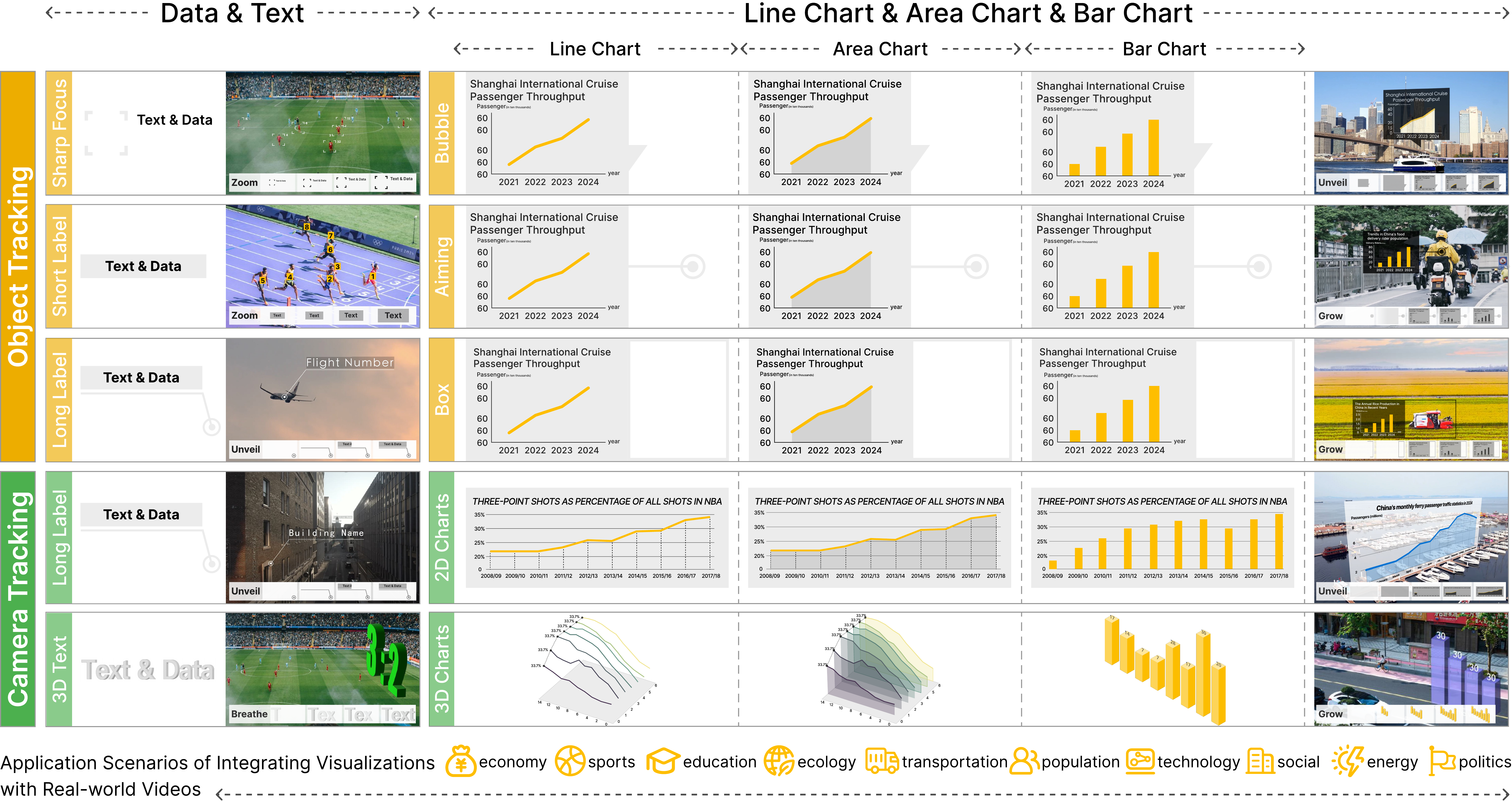}
    \caption{ Design space for visualizations suitable for embedding in video, which illustrates five representative visualization forms combined with two motion anchoring strategies. It reveals common design patterns for how visualizations are structured and integrated within dynamic video scenes across diverse real-world applications.}
    \label{fig:design-space}
\end{figure*}

\hy{
Furthermore, we identified several recurring design patterns in how visualizations are embedded within real-world video content. These patterns are summarized as follows.
\textbf{(1) Legibility.} Background panels are widely adopted to enhance readability (130/502), with a minority employing semi-transparent panels to maintain visual context (21/502). Gridlines are used selectively: horizontal guides appear more frequently (109) than vertical ones (43), while a notable number of charts omit gridlines entirely (64).
\textbf{(2) Referential Linking.} When visualizations correspond to specific objects in the scene, designers often employ visual linking techniques (e.g., arrows, lines, or bounding boxes) to establish clear referential ties (68). The choice of pointer type tends to adapt to scene complexity: in simple scenes with sparse motion, minimal markers like triangles often suffice; in cluttered environments with overlapping movements, bounding boxes or anchored pointers are preferred to unambiguously identify the target.
\textbf{(3) Simplicity.} Legends are rarely used (only 4 instances), reflecting a preference for minimalism in video formats. In contrast, essential structural elements such as axes (retained in 171 charts) and titles (present in 168 charts) are commonly preserved to maintain baseline interpretability.
\subsubsection{Design Space}
These findings highlight a set of implicit yet widely adopted design patterns for video-suited visualizations design, which we synthesize into a structured design space, as illustrated in Fig.~\ref{fig:design-space}. Notably, although we identified a broad spectrum of ten chart types across diverse application scenarios (as shown in Fig.~\ref{fig:chart}), we focused the design space on the five most commonly used chart types in the corpus to maximize generalizability.

The space is organized along three primary dimensions:
(i) \textbf{chart type}, covering five common forms, including data, text, line charts, bar charts, and area charts; 
(ii) \textbf{motion mode}, including object-tracking and camera-tracking embeddings; and 
(iii) \textbf{temporal behavior} shown at the bottom of each schematic. We describe each visual element by its on-screen duration and animation scheme.

Importantly, these dimensions are not independent; instead, they co-vary with production constraints and communicative intent in systematic ways. 
First, the choice of \textbf{motion mode} usually reflects what the visualization refers to. When the message is about a specific entity in the scene (e.g., a person, a vehicle, or a building), creators tend to use object-tracking so the overlay stays visually attached to the target. In these cases, the visualization is often kept compact (short text/data, small charts) and paired with explicit cues such as arrows or boxes. In contrast, when the visualization summarizes the scene or provides background context (e.g., overall trends or comparisons), creators more often use \textit{camera-tracking} overlays. This allows larger chart panels and reduces the risk that the overlay jitters or drifts with local object motion.
Second, \textbf{temporal behavior} is largely driven by attention and editing rhythm. Embedded charts rarely stay static on screen. Instead, they appear when the narration or shot content calls for data, remain long enough to be read, and then disappear to return attention to the footage. As a result, the most common animations are subtle and functional (e.g., fade-in/out or simple unveil), because aggressive animation competes with motion in the video and makes reading harder.

Finally, these observations explain why a design space is useful for system building. In professional workflows, creators repeatedly adjust the same small set of things: what the overlay is anchored to, how strictly it tracks, where it sits to avoid covering important content, how strong the background panel should be for legibility, and when it enters/exits. The design space guided our system's chart library design by turning the most frequently adjusted parameters into explicit controls and offering presets for common embedded chart styles.


}

\hy{
\subsection{Interaction}
To support efficient authoring and reliable synchronization, \oursName\ provides an integrated workflow that covers data binding, spatial anchoring, and temporal configuration. Key interaction techniques include:

\textbf{Upload and Data Mapping.} Users upload video footage and datasets via drag-and-drop or file selection. When timestamps are available, the system aligns data records with the video timeline. Users map data fields to visual channels by dragging them onto chart components, and the preview updates immediately to reflect mapping changes.

\textbf{One-click Tracking Anchor.} Users specify a tracking anchor by simply clicking a pixel in the video frame. Based on the selected motion mode, the system either (1) attaches the visualization to a moving target via Object Tracking or (2) keeps the chart anchored to the static scene via Camera Tracking in world coordinates.

\textbf{3D Adjustment.} Users pan, zoom, and rotate the chart in 3D to fine-tune placement: dragging pans, scrolling zooms, and right-dragging rotates the view. These operations work together with anchoring to achieve stable alignment and perspective-consistent embedding.

\textbf{Real-time Preview.} \oursName\ provides continuous, in-context preview throughout authoring. While editing, users can immediately see changes to chart styling and 3D viewpoint directly on the video. After rendering, users can play back the result to preview the chart’s motion in real time and verify alignment and visual coherence over time.

\textbf{Multi-chart Editing.} Users manage when each visualization appears using the timeline. Multiple charts are organized as separate tracks and can be added or deleted with simple timeline operations. Users adjust a chart segment’s entry, exit, and duration by dragging its endpoints, and selecting a segment brings that chart into focus for independent editing of its data mapping, style, and placement.}
\section{Evaluation}

\hy{We evaluated \textit{\oursName} through (1) two quantitative experiments that assess the precision of our two tracking pipelines and (2) interviews with domain experts that assess system usability. }

\subsection{Evaluation of Two Tracking Pipelines}
\hy{
Our system relies on two tracking pipelines: \textit{camera tracking} and \textit{object tracking}. The \textit{camera tracking} pipeline consists of~(A) depth estimation followed by~(B1) camera pose estimation. The \textit{object tracking} pipeline consists of~(A) depth estimation followed by~(B2) object trajectory estimation. The two pipelines share the same depth estimation stage. However, to the best of our knowledge, there's no open-source system that provides an end-to-end baseline for motion-synchronized chart embedding. We therefore construct baselines using a component substitution strategy: we replace individual stages with representative models identified in our literature review and evaluate the resulting combinations under identical experimental settings. This strategy allows us to quantify the tracking precision of two pipelines.}
\hy{\subsubsection{Precision of the Camera Tracking Pipeline.}
\par{\textbf{Experimental Setup.}}
The camera tracking pipeline comprises two core stages: depth estimation and camera pose estimation. In \oursName, the camera tracking pipeline is implemented as \textit{Metric3D + Open3D RGBD odometry}~\cite{Zhou2018,park2017colored}. To validate the effectiveness of our pipeline, we construct a baseline by substituting the depth estimation component, yielding \textit{ZoeDepth + RGBD odometry} (baseline). We choose this baseline because Open3D RGB-D odometry is a frequently used and stable open-source camera pose estimator, making it a reliable and fixed module. \textit{ZoeDepth}~\cite{bhat2023zoedepth} is another top-performing depth estimator identified in our literature review. The goal of our experiment is to compare whether our depth estimation model~(Metric3D) achieves better performance than ZoeDepth when paired with the same tracking module.

We evaluate on the subset of DriveTrack~\cite{koppula2024tapvid} (18 sequences, 1,806 frames) with ground-truth camera poses, where camera motion is sufficiently pronounced, and scenes include dynamic foreground objects. These conditions closely resemble the real-world videos in which our system embeds charts. For each frame, the predicted depth map is fed into RGBD odometry to estimate camera poses.

\par{\textbf{Evaluation Metric.}}
We report rotation and translation errors between predicted poses $(R_{\text{pred}}, t_{\text{pred}})$ and ground truth $(R_{\text{gt}}, t_{\text{gt}})$. Rotation error is computed from the relative rotation $\Delta R = R_{\text{gt}}^{-1}R_{\text{pred}}$ as the corresponding angular difference, and translation error is the Euclidean distance $\lVert t_{\text{gt}} - t_{\text{pred}} \rVert$.

\par{\textbf{Results and Analysis.}}
Table~\ref{tab:static-anchoring} shows that our pipeline yields lower rotation and translation errors than the baseline (mean and median). Since both variants use the same odometry backend and camera intrinsics, the improvement is attributable to the depth backbone, indicating that Metric3D provides depth estimates that better support downstream pose recovery in our pipeline.
}

\begin{table}[htbp]
  \centering
  \caption{Comparison on camera tracking.
 (Camera pose estimation errors)}
  \label{tab:static-anchoring}
  \resizebox{\linewidth}{!}{
  \begin{tabular}{lcccc}
    \toprule
    Method & \multicolumn{2}{c}{Rotation Error (degrees)} & \multicolumn{2}{c}{Translation Error (m)} \\
    \cmidrule(lr){2-3} \cmidrule(lr){4-5}
           & Mean ↓ & Median ↓ & Mean ↓ & Median ↓ \\
    \midrule
    \textbf{Metric3D + RGBD odometry (ours)} & \textbf{8.641} & \textbf{8.851} & \textbf{6.518} & \textbf{6.432} \\
    ZoeDepth + RGBD odometry                 & 10.544         & 11.061         & 7.598          & 7.610 \\
    \bottomrule
  \end{tabular}}
\end{table}

\hy{

\subsubsection{Precision of the Object Tracking Pipeline.}
\par{\textbf{Experimental Setup.}}
Object tracking comprises two stages: depth estimation and object trajectory estimation. In \oursName, we implement this pipeline as \textit{Metric3D + BootsTAP}, which is not the only reasonable choice. During our literature review, we also identified strong alternatives for each stage, including \textit{ZoeDepth}\cite{bhat2023zoedepth} for depth estimation and \textit{TAPIR}\cite{doersch2023tapir} for object trajectory estimation. To validate that our chosen pipeline remains the best-performing configuration under the same experimental conditions, we evaluate three additional baseline combinations: \textit{Metric3D + TAPIR}, \textit{ZoeDepth + TAPIR}, and \textit{ZoeDepth + BootsTAP}.

We performed this evaluation on a subset of the \textit{PStudio} dataset~\cite{joo2015panoptic}, which contains high-fidelity multiview recordings of social interaction scenes. We selected 36 video sequences (10,800 frames total) that include clearly visible dynamic objects undergoing complex 3D motion, while the camera remains mostly stationary. This setup allows us to isolate the quality of object trajectory reconstruction from the effects of camera movement. 
}

\hy{
\par{\textbf{Evaluation Metric.}}
We adopted the \textit{Average 3D Point Distance} (APD\textsubscript{3D}) metric~\cite{koppula2024tapvid}, to evaluate the precision of 3D point tracking. APD\textsubscript{3D} measures the fraction of predicted 3D points that fall within a given Euclidean distance threshold of the corresponding ground-truth points, averaged over time and across tracks. Higher values indicate more accurate and temporally consistent 3D trajectories.}

\hy{
\par{\textbf{Results and Analysis.}}
As shown in Table~\ref{tab:dynamic-tracking}, pipelines using \textit{BootsTAP} consistently outperform those using \textit{TAPIR} under the same depth backbone, indicating that BootsTAP provides more reliable 2D keypoint trajectories for downstream 3D reconstruction.

Although \textit{ZoeDepth + BootsTAP} achieves the highest APD\textsubscript{3D} score, we adopt \textit{Metric3D + BootsTAP} in \oursName for better alignment with our camera tracking pipeline. Because both variants use the same 2D tracker (\textit{BootsTAP}), the APD\textsubscript{3D} gap primarily reflects the choice of depth backbone. With regard to depth estimation, our objective is not metrically exact 3D trajectory recovery, but perspective-consistent chart embeddings that preserve spatial coherence between foreground and background elements. Therefore, moderate depth errors are acceptable, provided that the overall depth structure and recovered trajectories remain stable.
}

\begin{table}[htbp]
  \centering
  \caption{Comparison on object tracking.
 (Average 3D point distance (APD$_\text{3D}$)).}
  \label{tab:dynamic-tracking}
  
  \begin{tabular}{lc}
    \toprule
    Method & APD$_\text{3D}$ (↑) \\
    \midrule
     Metric3D + BootsTAP  (ours)           & 0.3424 \\
     Metric3D + TAPIR                     & 0.3416 \\
    ZoeDepth + BootsTAP                  & \textbf{0.3609} \\
    ZoeDepth + TAPIR                       & 0.3595 \\
    \bottomrule
  \end{tabular}
\end{table}

\begin{figure*}

    \centering
    \includegraphics[width=0.8\linewidth]{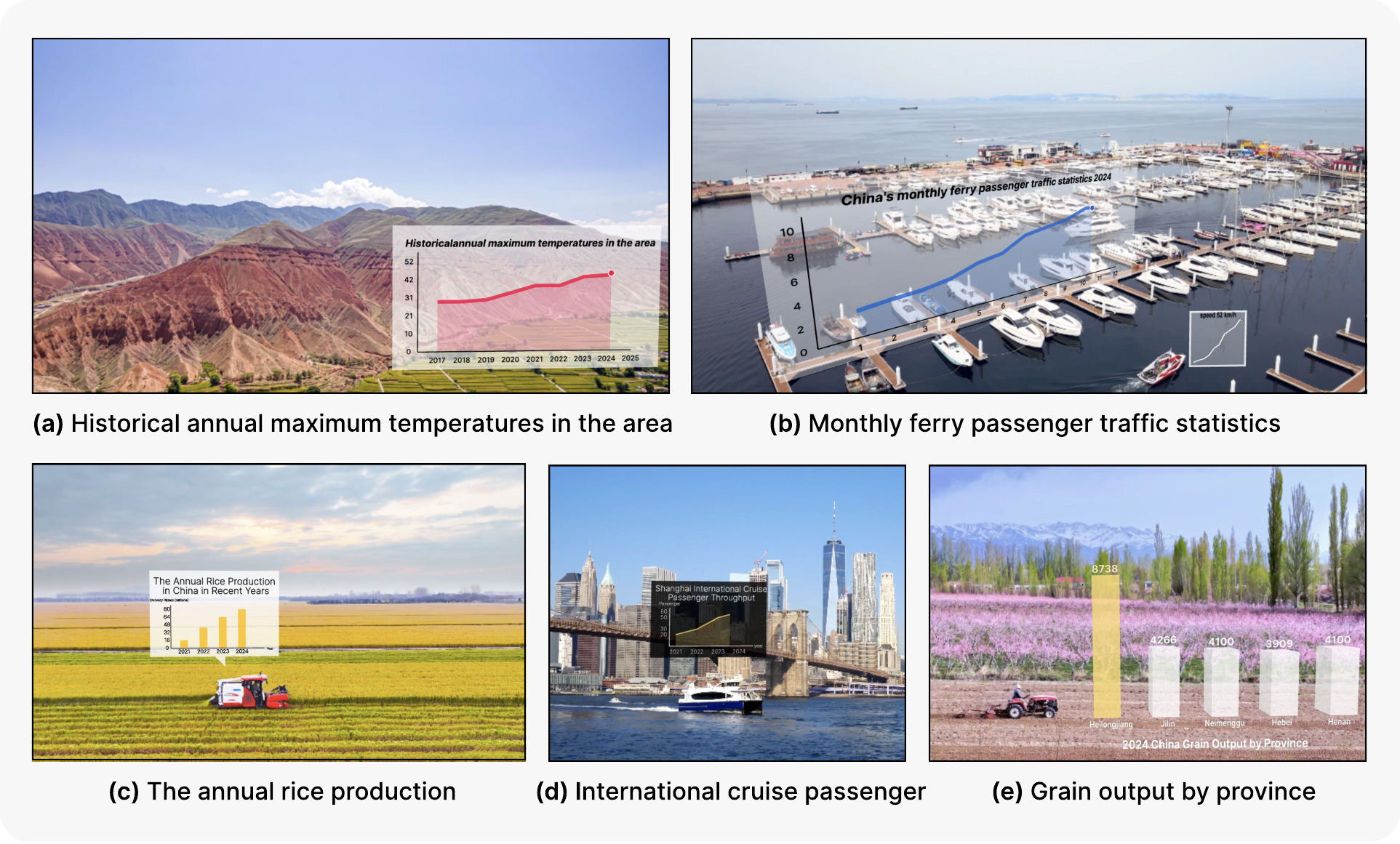}
    \caption{Five example videos with embedded visualizations generated using \textit{\oursName}}
    \label{fig:case}
\end{figure*}
\subsection{Expert Interview}
\hy{\par{\textbf{Participants.}} To further evaluate the usability of \textit{\oursName}, we conducted semi-structured interviews with five data journalists (E1–E5; Male=2, Female=3; Age: 32-45). Two were from our collaborating media organization, while the remaining three were recruited from other newsrooms to broaden perspectives and reduce organizational bias. Participants had, on average, over ten years of video-editing experience and were familiar with chart-embedded videos. All participants provided informed consent before the study.}
\par{\textbf{Procedure.}} The interviews were performed via an online meeting
system. Each interview started with a 10-minute introduction about the system. Participants were first introduced to our system through a brief tutorial delivered by the authors and were given two tasks:

\textit{Task 1 - Goal-oriented (10 minutes).} Participants are shown an example video that they have to reproduce. The result should consist of a bar chart that remains fixed in the field and an area chart that moves along with the vehicle.

\textit{Task 2 - Open-ended (30 minutes).} Participants are asked to create two novel data videos by choosing among 10 short video clips (10" duration on average).

\hy{\textbf{Measures.} After the tasks, experts participated in a semi-structured interview. Each session lasted approximately 60 minutes. We gathered two data sources: (i) The completed videos, (ii) audio recordings of think-aloud protocols and interviews. Transcripts and observation notes were analyzed using a reflexive thematic analysis. Three authors independently performed open coding to generate candidate codes, reconciled discrepancies through discussion, and iteratively refined a shared codebook until reaching consensus on higher-level themes. }


\par{\textbf{Result.}}
Fig.~\ref{fig:case} presents sample videos generated by expert users using our system. In follow-up interviews, the experts offered a range of insightful comments, which are summarized as follows:

\textit{\underline{Fast Crafting.}} All experts agreed that our system was highly effective in supporting both camera tracking and object tracking. In particular, E1 commented that \textit{``This system can accomplish the work of several hours, or even days, in just two or three minutes.''} E2 mentioned that \textit{``This system allows me to see the real-time effect of chart edits in the video... I no longer need to repeatedly switch between project files, which used to waste a lot of time."} E5 also noted that \textit{``The speed of result generation is so \textbf{fast} that I can quickly create multiple demos to observe where the chart embedding works best in the video."}

\textit{\underline{Accurate Results.}} Most experts were very satisfied with our results, both for camera and object tracking (E1, E2, E4, E5). E1 commented, \textit{``The results are directly \textbf{usable}.''} E2 suggested that \textit{``The object tracking techhnique \textbf{amazed} me; it not only takes the camera pose into account but also integrates the movement of the objects.'' }E5 mentioned, \textit{``This result is more accurate than manual alignment.''}

\textit{\underline{A Professional and Elegant System.}} All the experts were satisfied with the system. Additionally, all of them mentioned that the system was very user-friendly and that they would recommend it to others. Specifically, they appreciated how seamlessly the system integrates chart creation and embedding into videos, accomplishing highly professional tasks (E1, 2, 5). E1 mentioned that \textit{``This system can complete highly \textbf{professional tasks} without requiring extensive skills or knowledge."} E5 noted that \textit{``the chart can be translated and rotated in \textbf{3D}, and it integrates seamlessly with the video.''} They were also satisfied with the chart creation aspect. E4 noted that \textit{``the data mapping feature eliminated the need for selecting data and creating visualizations in 3D software."} E5 commented that \textit{``The chart creation process is very \textbf{user-friendly} for me. There are many \textbf{expressive} templates available for me to use ... Previously, I had to spend a lot of time on creating visualizations."}

\section{LIMITATIONS AND FUTURE WORK}
\hy{
While the evaluation results indicate \oursName is promising to help users embed and synchronize charts with video scenes, the system still has several
limitations that were found during the implementation or mentioned by the participants during interviews. We hope to guide potential future research directions by pointing out these limitations.}

\textit{\underline{Context-Aware Visualization Generation.}} Several experts (E1, E2, E4) expressed the need for a more intelligent system capable of understanding both dataset and video semantics. For instance, the system could automatically suggest visualization color schemes aligned with the video or incorporate domain-specific visual metaphors, such as rendering bars in a bar chart as crops in agricultural contexts. It could also generate appropriate chart titles based on the video narrative and the underlying data. Future work may explore fine-tuning Large Language Models (LLMs) to support automatic topic analysis, enabling the system to generate context-aware visualizations with optimized design elements, enhancing the coherence and quality of data-driven videos.


\textit{\underline{Tracking Editing.}} 
Currently, the system automatically generates motion trajectories for embedded charts based on anchor points. A promising extension is to introduce a visual trajectory editor that renders the underlying coordinate array as editable curves within a canvas or SVG-based interface. Leveraging techniques such as spline interpolation, draggable control points, and direct path manipulation, users could intuitively modify the motion path. Such an interface would offer frame-level control over trajectory shapes, allowing insertion, deletion, and adjustment of keypoints. This would significantly enhance the flexibility and expressiveness of the authoring workflow.

\hy{\textit{\underline{Semantic Motion Analysis.}} 
While our system employs robust pixel‑level tracking to ensure precise geometric synchronization, it lacks the ability to handle occlusions effectively~(E1,E2,E3). This limitation stems from its reliance on low‑level visual features without integrating higher‑level semantic reasoning. We regard pixel‑level tracking as a fundamental requirement for stable embedding, as it generalizes well across diverse video types without being constrained by predefined object categories. Introducing semantic recognition could effectively address motion‑occlusion challenges, enabling the system to maintain tracking continuity even when visual features are temporarily obscured.
}

\section{Conclusion}
In this paper, we introduced \textit{\oursName}, an integrated system for authoring, embedding, and synchronizing data visualizations within dynamic video content. By combining computer vision algorithms for camera and object tracking, we enable synchronization between visualizations and video motion. In addition, the system incorporates an interactive workflow to support easy and efficient authoring of videos with visualizations embedded.
Quantitative evaluations demonstrate the effectiveness of our algorithm pipeline
in recovering spatially consistent motion trajectories. Additionally, interviews with domain experts confirm the system’s usability, expressiveness, and potential for professional media production. The evaluation showed the power of \oursName\ system and revealed several limitations of the current system, which will be addressed in the future.
\section*{ACKNOWLEDGMENTS}
Nan Cao is the corresponding author. This work was supported by the National Key Research and Development Program of China (2023YFB3107100). We would like to thank all the reviewers for their valuable feedback.

\bibliographystyle{IEEEtran}
\bibliography{sections/reference}

\begin{IEEEbiography}[{\includegraphics[width=1in,clip,keepaspectratio]{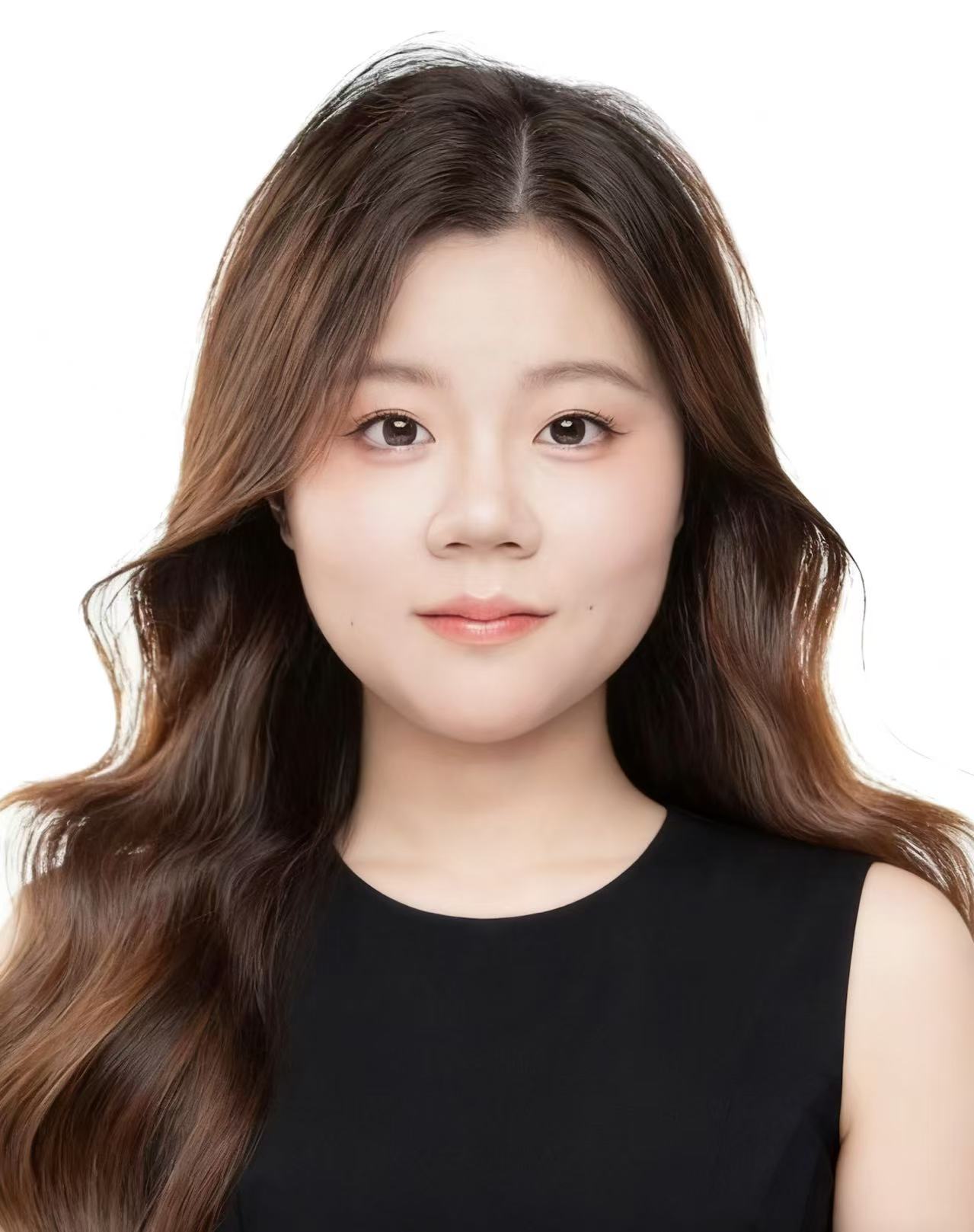}}]{Yi He} received her B.Eng degree from School of Digital Media \& Design Arts, Beijing University of Posts and Telecommunications.  She is currently working toward her Ph.D. degree as part of the Intelligent Big Data Visualization (iDV$^x$) Lab, Tongji
University. Her research interests include data
visualization and artificial intelligence.
\end{IEEEbiography}
\vspace{-32pt}
\begin{IEEEbiography}[{\includegraphics[width=1in,clip,keepaspectratio]{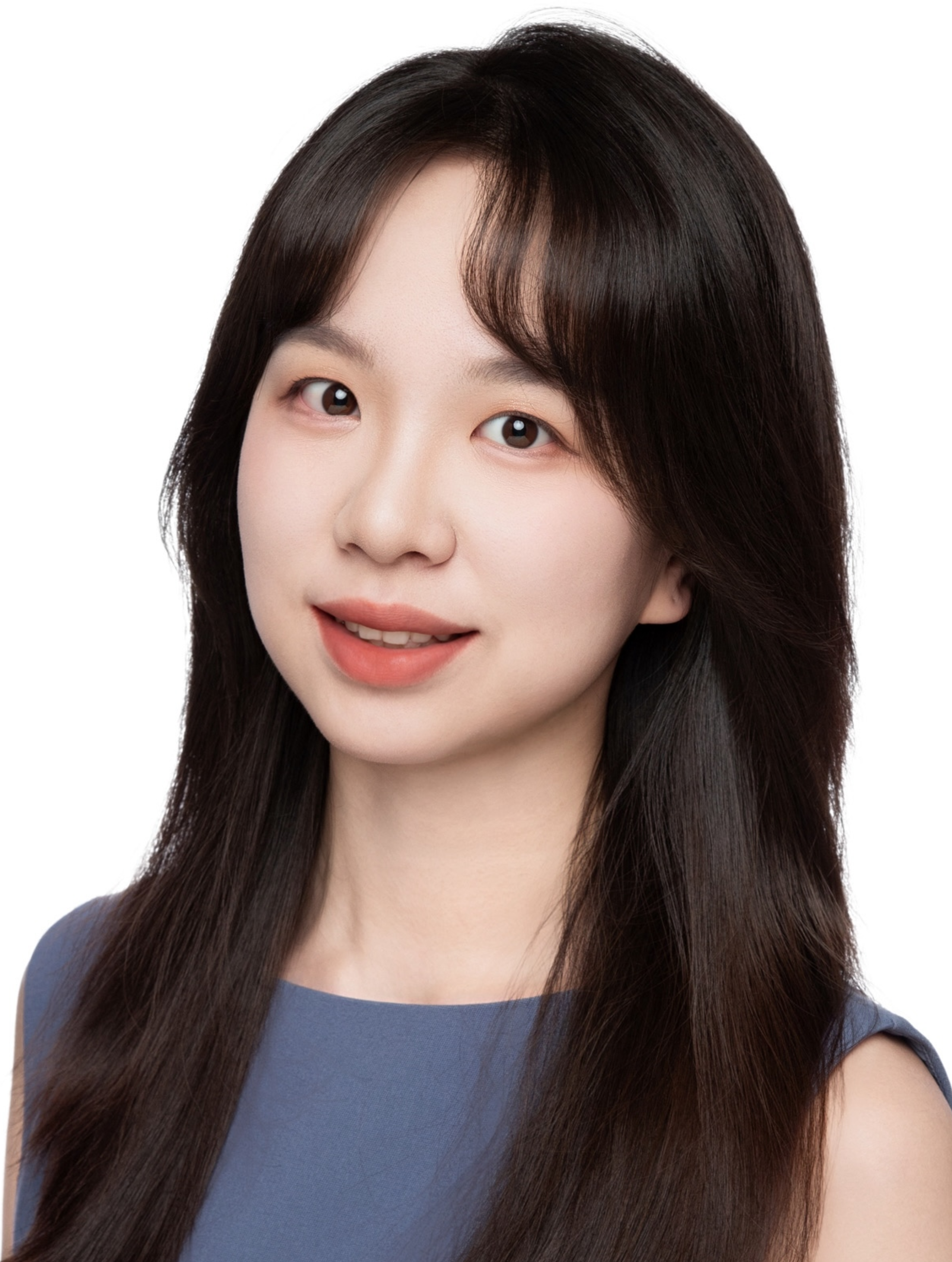}}]{Yuqi Liu} received her B.Eng degree from the School of Computer Science and Technology, Shanghai University of Computer Engineering and Science. She is currently working toward her Master’s degree as part of the Intelligent Big Data Visualization (iDV$^x$) Lab at Tongji University. Her research interests include data visualization and artificial intelligence.
\end{IEEEbiography}
\vspace{-32pt}
\begin{IEEEbiography}[{\includegraphics[width=1in,clip,keepaspectratio]{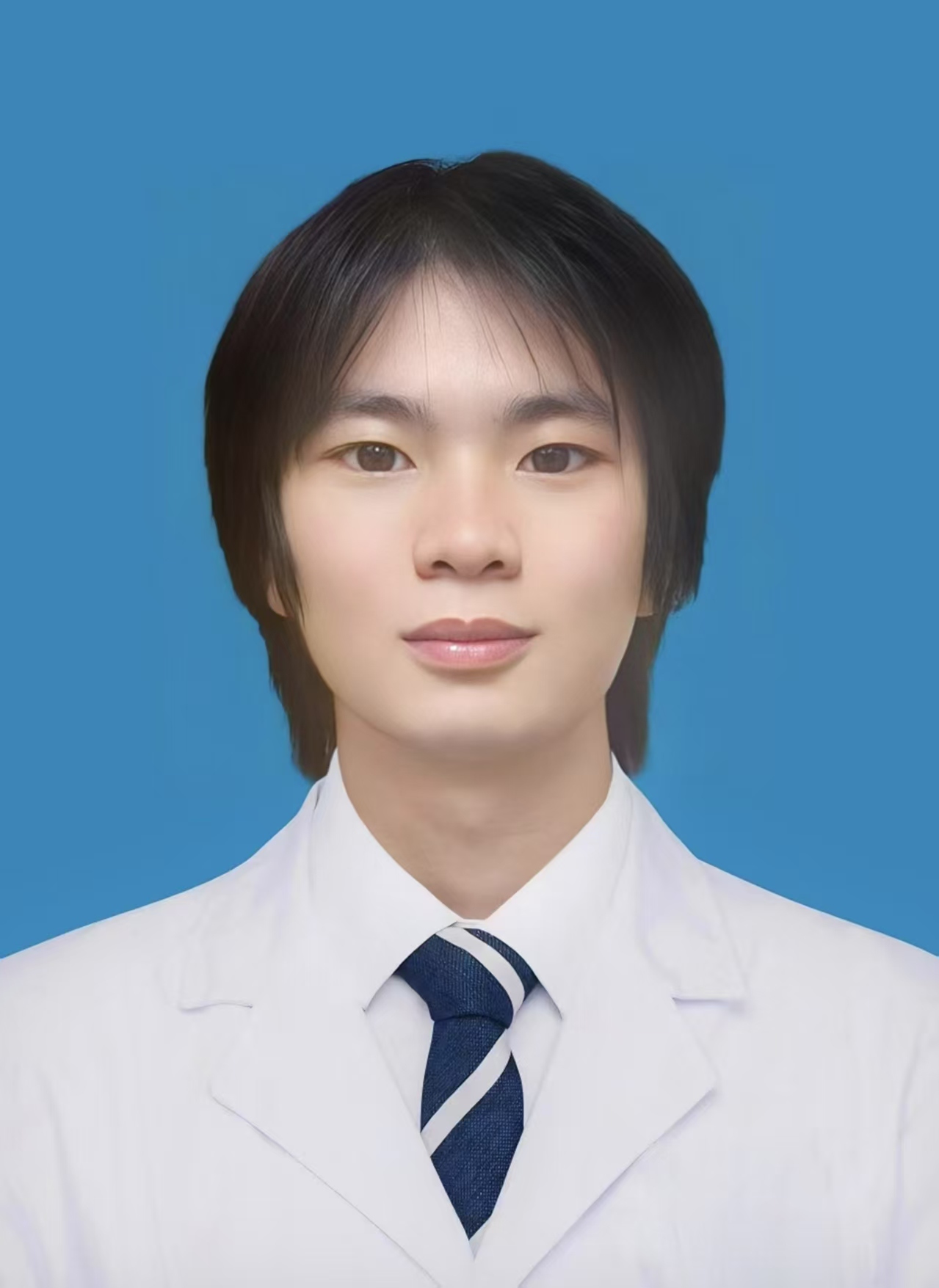}}]{Chenpu Li} is pursuing his B.Eng dual degree in Visual Communication Design and Artificial Intelligence at the College of Design and Innovation, Tongji University. His cur-rent academic focus spans Human-Computer Interaction (CHI) and AI-Art integration.
\end{IEEEbiography}
\vspace{-32pt}
\begin{IEEEbiography}[{\includegraphics[width=1in,clip,keepaspectratio]{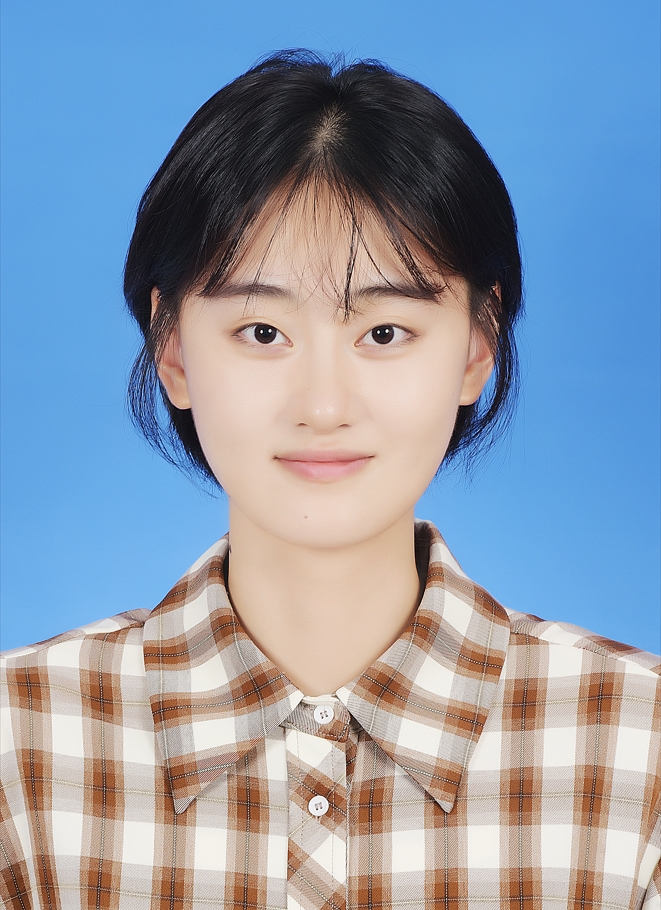}}]{Ruoyan Chen} received her B.Eng degree in Industrial Design from University of Science and Technology Beijing. She is currently pursuing her M.S. degree as part of the Intelligent Big Data Visualization (iDV*) Lab, Tongji University. Her research interests include data visualization and agent design.
\end{IEEEbiography}
\vspace{-32pt}
\begin{IEEEbiography}[{\includegraphics[width=1in,clip,keepaspectratio]{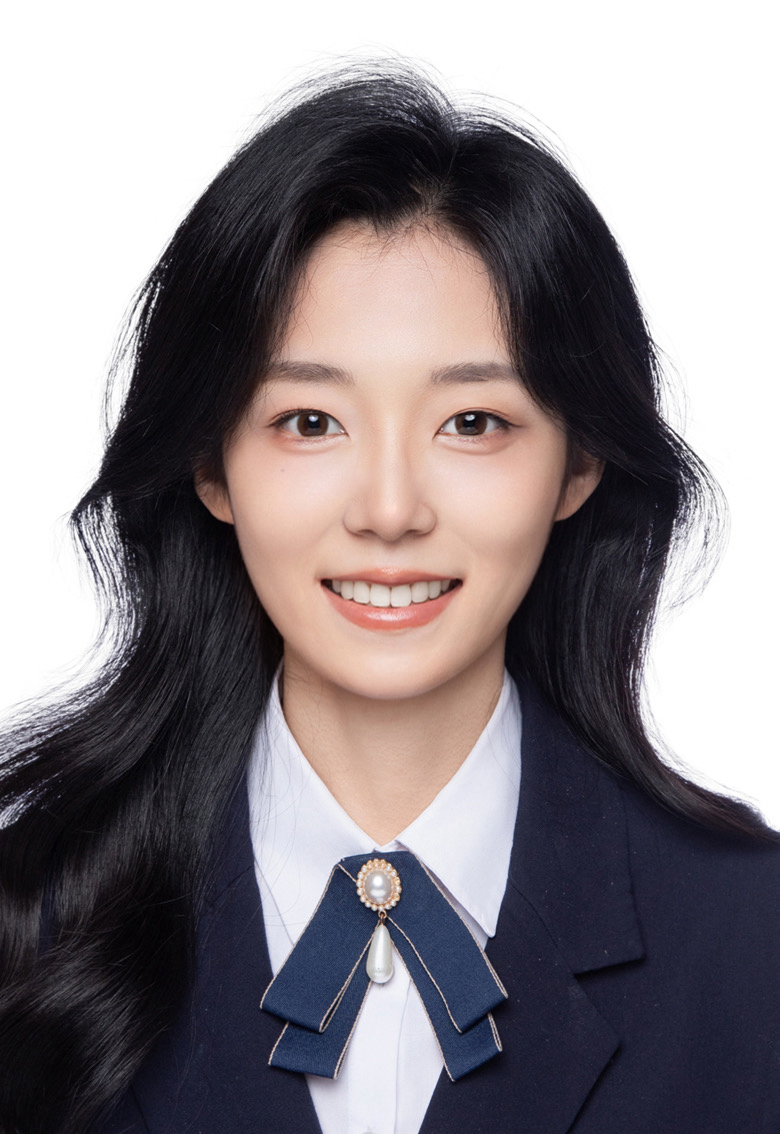}}]{Chuer Chen} received her MSc degree from the Department of Electrical and Computer Engineering, National University of Singapore in 2021. She is currently working toward her Ph.D. degree as part of the Intelligent Big Data Visualization (iDVx) Lab, Tongji University. Her research interests include information visualization and intelligent design.
\end{IEEEbiography}
\vspace{-32pt}
\begin{IEEEbiography}[{\includegraphics[width=1in,clip,keepaspectratio]{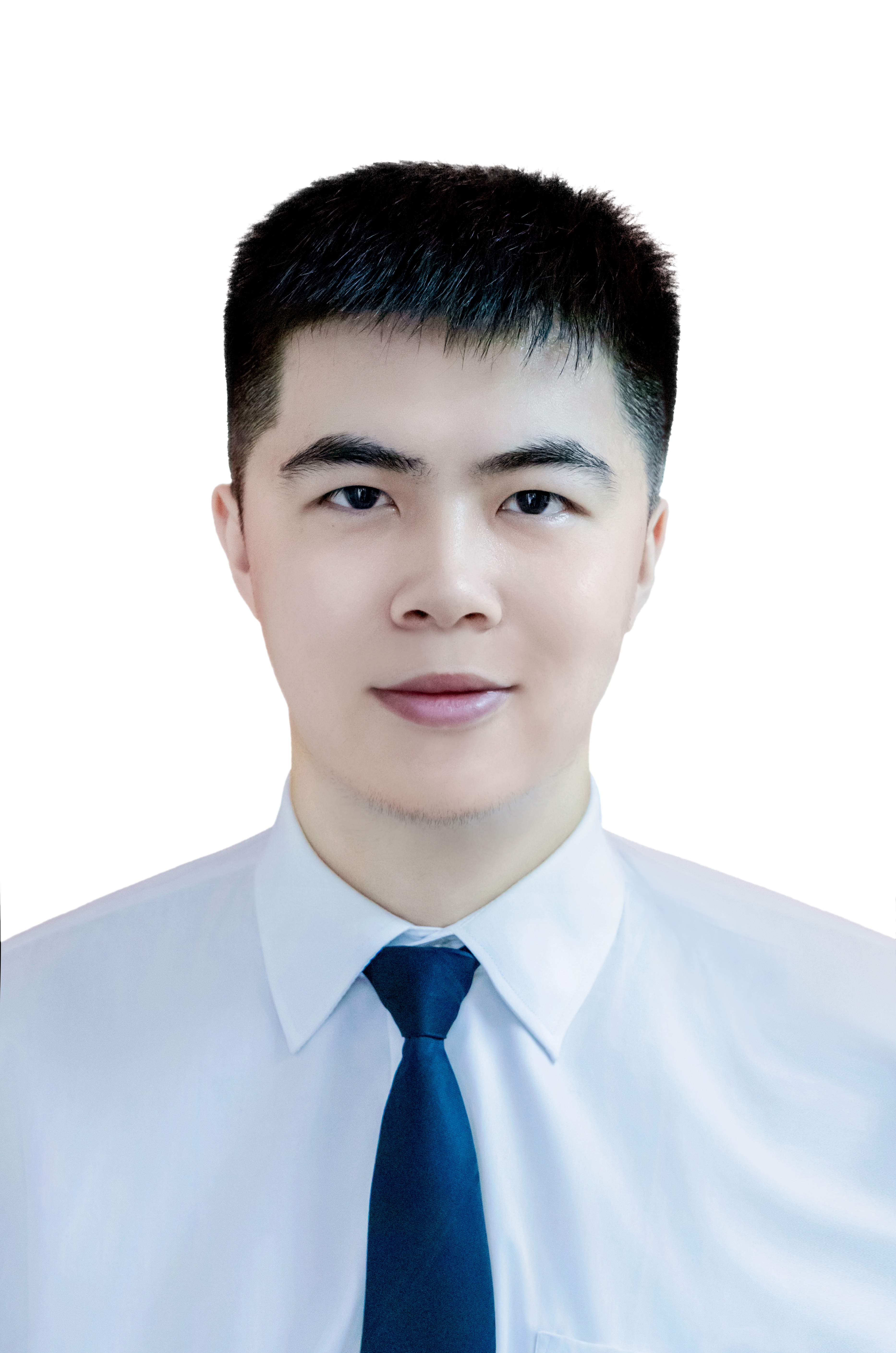}}]{Shengqi Dang} received his B.Sc. in Mathematics from Tongji University and is currently pursuing a Ph.D. degree at the Intelligent Big Data Visualization (iDVx) Lab, Tongji University. His research interests include of artificial intelligence and computer graphics.
\end{IEEEbiography}
\vspace{-32pt}
\begin{IEEEbiography}[{\includegraphics[width=1in,height=1.25in,clip,keepaspectratio]{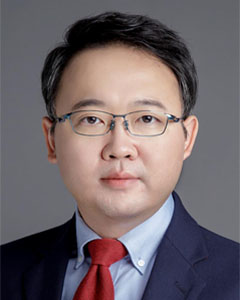}}]{Nan Cao} received his Ph.D. degree in Computer Science and Engineering from the Hong Kong University of Science and Technology (HKUST), Hong Kong, China in 2012. He is currently a professor at Tongji University and the Assistant Dean of the Tongji College of Design and Innovation. He also directs the Tongji Intelligent Big Data Visualization Lab (iDV$^x$ Lab) and conducts interdisciplinary research across multiple fields, including data visualization, human computer interaction, machine learning, and data mining. He was a research staff member at the IBM T.J. Watson Research Center, New York, NY, USA before joining the Tongji faculty in 2016.
\end{IEEEbiography}

\end{document}